\documentclass{article}

\usepackage{graphicx}
\graphicspath{%
    {converted_graphics/}
    {/}
}
\begin{document}

\title{Effect of Gravitational Frame Dragging on Orbiting Qubits}         
\author{Marco Lanzagorta\thanks{\texttt{marco.lanzagorta at nrl.navy.mil}} \\Code 5543\\US Naval Research Laboratory}        
\date{}          
\maketitle

\begin{abstract}
In this paper  we discuss the effect of gravitational frame dragging on orbiting qubits. In particular, we consider 
the Kerr spacetime geometry and spin-$\frac{1}{2}$ qubits moving in an equatorial radial fall with zero angular momentum and 
equatorial circular orbits. We ignore the ${\cal O}(\hbar)$ order effects due to spin-curvature coupling, which allows us to consider the
motion of the spin-$\frac{1}{2}$ particles 
as Kerr geometry geodesics. We derive analytical expressions for the infinitesimal Wigner rotation and numerical
results for their integration across the length of the particle's trajectory. To this end, we consider the bounds on
the finite Wigner rotation imposed
by Penrose's cosmic censorship hypothesis. 
\end{abstract}

\section{Introduction} 

One of the major scientific thrusts from recent years has been to try to harness quantum phenomena
to dramatically increase the performance of a wide variety of classical information processing devices.
In particular, it is generally accepted that quantum computers and communication systems promise
to revolutionize our information infrastructure.

With the prospect of satellite based quantum communications, it is necessary to understand the dynamics
of qubits in such an operational environment.
To this end, a considerable amount of work has been done to understand how entanglement and quantum information
are affected by Lorentz transformations in the context of special relativity \cite{alsing, adami, adami2, terno, soo}. 
Further work has been reported on the effects of spacetime curvature on 
quantum information \cite{ueda, fuentes2, fuentes, alsing3, alsing2, alsing1}. 
Most of this research is based on
the analysis of spinor and vector quantum fields coupled to gravity in the context of General 
Relativity \cite{boulware1, boulware2, davis, parker}.

The specific analysis of qubits in stationary axisymmetric spacetimes produced by rotating bodies, such as the one
represented by the Kerr metric, 
has not been considered in the literature. These spacetimes are important because of the Lense-Thirring effect, which 
does not have a classical counterpart.
Here, spacetime is {\it dragged} by the rotating sources of the gravitational field \cite{chandra, hobson}. 
The dynamics of spin-$\frac{1}{2}$ particles in the gravitational field of rotating bodies has been studied in considerable 
detail \cite{kumar, lalak, bani, nicolas}. 
 
In this paper, 
we discuss the effect of gravitational frame dragging on orbiting qubits. In Section 2 we offer a brief overview of
the Wigner rotations induced on qubit states due to spacetime curvature \cite{ueda}. The structure of Kerr spacetime, 
geodesics, and associated tetrad fields and connection 1-forms are presented in Section 3. The analysis of qubits in
Kerr spacetime is presented in Section 4. We restrict our analysis to the case of an equatorial radial fall with zero
angular momentum and equatorial circular orbits.

\section{Wigner Rotations in Curved Spacetime}

General coordinate transformations in general relativity are described through the group $GLR(4)$ made of
all real regular $4\times 4$ matrices. However, it is known that $GLR(4)$ does not have a spinor 
representation \cite{cartan}. 
Therefore, the best way to study the dynamics of spin-$\frac{1}{2}$ particles in gravitational fields is through  
the use of local inertial frames defined at each point of spacetime \cite{ueda}. These local inertial frames 
are defined through a tetrad field $e^a_{~\mu}(x)$, which is a set of four linearly independent coordinate
4-vector fields \cite{zakharov}. 
In what follows, Latin indices $a,b,c...$ refer to 
coordinates in the local inertial frame, while Greek indices $\mu,\nu...$ correspond to the general coordinate
system. 

The Minkowsky metric $\eta^{ab}$ in the local inertial frame and 
the spacetime metric tensor $g^{\mu\nu}(x)$ are related through the tetrad field:
\begin{eqnarray}
g^{\mu\nu}(x) &=&e_a^{~\mu}(x)~e_b^{\nu}(x)~ \eta^{ab} \\\nonumber
\eta^{ab}&=&e^a_{~\mu}(x)~e^b_{\nu}(x)~ g^{\mu\nu}(x)\nonumber
\end{eqnarray} 
In a similar manner, the momentum in the local inertial frame is related to the momentum in the general coordinate system by:
\begin{equation}
p^a(x)=p^\mu(x) ~e^a_{~\mu}(x)
\end{equation}

Then, any change in the momentum in the local inertial frame can be expressed as a combination
of changes due to (1) external forces other than gravity $\delta p^\mu(x)$, and (2) spacetime
geometry effects (gravity) $\delta e^a_{~\mu}(x)$:
\begin{equation}
\delta p^a(x)=\delta p^\mu(x) ~e^a_{~\mu}(x)+p^\mu(x)~\delta e^a_{~\mu}(x)
\end{equation}
where:
\begin{eqnarray}
\delta p^\mu(x) &=& m ~a^\mu(x)~ d\tau \\\nonumber
\delta e^a_{~\mu}(x) &=& -u^\nu(x)~\omega^{~a}_{\nu~b}(x) ~ e^b_{~\mu}(x) ~d\tau
\end{eqnarray} 
and the connection 1-forms are defined by:
\begin{equation}
\omega^{~a}_{\nu~b}(x) =e^a_{~\nu}(x)~\nabla_\mu e_b^{~\nu}(x)
\end{equation}
and $\nabla_\mu$ is the covariant derivative. Notice that $a^\mu(x)$ is the 4-acceleration due to 
external forces other than gravity.

As the particle moves, its momentum in the local inertial frame will transform under a  local Lorentz 
transformation:
\begin{equation}
p^a(x)=\Lambda^a_{~b}(x)~p^b(x)
\end{equation} 
where:
\begin{equation}
\Lambda^a_{~b}(x)=\delta^a_{~b}+\lambda^a_{~b}(x)d\tau
\end{equation} 
In the absence of external forces, $a^\mu(x)=0$ and the 
Lorentz transformation in the local inertial frame is simply given by:
\begin{equation}
\lambda^a_{~b}(x)=-u^\nu(x)~\omega^{~a}_{\nu~b}(x) 
\end{equation} 
where $u^\mu(x)$ is the 4-velocity of the particle.

The associated infinitesimal Wigner rotation that affects the particle's spin is given by:
\begin{equation}
W^a_{~b}(x)=\delta^a_{~b}+\vartheta^a_{~b}(x)~d\tau
\end{equation} 
where:
\begin{equation}
\vartheta^a_{~b}(x)=\lambda^a_{~b}(x)+\frac{\lambda^a_{~0}(x)~u_b(x)-\lambda_{b0}(x) ~u^a(x)}
{u^0(x)+1}
\end{equation}
and its spin-$\frac{1}{2}$ representation is:
\begin{equation}
D_{\sigma'\sigma}^{(1/2)}(W(x))=I+\frac{i}{2}\left(\vartheta_{23}(x)\sigma_x  + \vartheta_{31}(x)\sigma_y   + 
\vartheta_{12}(x)\sigma_z   \right)d\tau
\end{equation}
where $\sigma_{x,y,z}$ are tha Pauli matrices \cite{ueda}.

Therefore, the Wigner rotation for a particle that moves over a finite proper time interval is:
\begin{equation}
W^a_{~b}(x_f;x_i)~=~:\exp \left( \int_{\tau_i}^{\tau_f} \vartheta^a_{~b}(x(\tau))~d\tau\right) :
\end{equation} 
where $::$ indicates a time ordered product and the integral is taken along the path of the particle.

\section{The Kerr Geometry}

The Kerr metric is a solution to Einstein's field equations that represents a stationary axisymmetric 
spacetime produced by a rotating object of mass $M$ and angular momentum $Ma$ \cite{chandra, hobson}. 
This solution can be proved to be unique if one demands that:
(1) the spacetime tends to the Minkowsky form as $r\rightarrow\infty$ and (2) the geometry is non-singular
outside of a smooth closed convex event horizon. Under these conditions 
the Kerr metric is expressed in terms of the Boyer-Lindquist coordinates $(t,r,\theta,\phi)$ in Planck units 
$(G=1,c=1)$ as:
\begin{equation}
dS^2=-\frac{\rho^2\Delta}{\Sigma^2}dt^2+\frac{\Sigma^2\sin^2\theta}{\rho^2}\left( d\phi - \omega dt \right)^2
+\frac{\rho^2}{\Delta}dr^2+\rho^2d\theta^2
\end{equation}
where:
\begin{eqnarray}
\rho^2&=&r^2+a^2\cos^2\theta\\\nonumber
\Delta&=&r^2-rr_s+a^2\\\nonumber
\Sigma^2&=&(r^2+a^2)^2-a^2\Delta\sin^2\theta\\\nonumber
r_s&=&2M\\\nonumber
\omega &=& \frac{r_sra}{\Sigma^2}\nonumber
\end{eqnarray}

The Schwarzschild metric is obtained in the no-rotation limit:
\begin{equation}
\lim_{a\rightarrow 0}dS^2=-f~dt^2+\frac{1}{f}~dr^2+r^2(d\theta^2+\sin^2\theta ~d\phi^2)
\end{equation} 
where:
\begin{equation}
f=1-\frac{r_s}{r}\\\nonumber
\end{equation}
As a consequence, all the results presented in the following sections also apply to the Schwarzschild metric by
setting the limit $a\rightarrow 0$.

\subsection{Structure of Kerr Spacetime}

The Kerr metric has two event horizons related to coordinate singularities \cite{carter}.
These occur on the surfaces:
\begin{equation}
r_{\pm}=\frac{r_s}{2}\pm\sqrt{\frac{r_s^2}{4}-a^2}
\end{equation}
Because of Penrose's cosmic censorship hypothesis, one has to demand the existence of 
event horizons that cover the true singularity, therefore avoiding 
troublesome naked singularities \cite{penrose}. In the Kerr metric case, the true singularity 
occurs at $\rho=0$. 
As a consequence, event horizons in Kerr spacetime exist if:
\begin{equation}
\frac{r_s^2}{4}>a^2~~~\Longrightarrow~~~~-0.5<\frac{a}{r_s}<0.5
\end{equation}

In addition, the Kerr geometry has two stationary limit surfaces $S^\pm$ defined by $g_{tt}(S^\pm)=0$. 
These are infinite redshift surfaces where the rotation of the compact object is so strong 
that any test particle is forced to rotate with the source, even if it has an arbitrarily large angular momentum \cite{hobson}. 
For the
case of the Kerr metric, these surfaces are found in:
\begin{equation}
S^{\pm}=\frac{r_s}{2}\pm\sqrt{\frac{r_s^2}{4}-a^2\cos\theta}
\end{equation}
As a consequence, the Kerr spacetime geometry has the following structure:
\begin{equation}
S^-\le r_- \le r_+ \le r_s \le S^+
\end{equation}

In the following discussions we will concentrate on qubits traveling in regions exterior to the stationary limit surface,
where the Boyer-Lindquist coordinates are well defined. It is possible, however, to analyze the 
interior of $S^+$ by using Eddington-Finkelstein coordinates, which provide an analytic continuation that extends
the range of validity of the equations \cite{carter}. This case will be analyzed in a future paper.

\subsection{Tetrads and 1-Forms}
It is convenient to choose the tetrad field that defines a local inertial frame in each point of Kerr spacetime as 
follows \cite{kumar}:
\begin{eqnarray}
e_0^{~t}(x)&=&\frac{1}{W}\\\nonumber
e_1^{~r}(x)&=&\frac{\sqrt{\Delta}}{\rho}\\\nonumber
e_2^{~\theta}(x)&=&\frac{1}{\rho}\\\nonumber
e_3^{~\phi}(x)&=&\frac{W}{\sqrt{\Delta \sin\theta}}\\\nonumber
e_3^{~t}(x)&=&\frac{a\sin\theta}{\sqrt{\Delta}} \left( W-\frac{1}{W}\right)
\end{eqnarray} 
where:
\begin{equation}
W=\sqrt{1-\frac{rr_s}{\rho^2}}
\end{equation}
Let us remark that the choice of tetrad field is somewhat arbitrary, and it could be defined in a different way \cite{bardeen}.
In the present case, the tetrad represents the local inertial frame of a ``hovering'' observer.

In the equatorial plane ($\theta=\pi/2$), the non-zero connection 1-forms associated to our choice of tetrad field are
given by \cite{kumar}:
\begin{eqnarray}
\omega_{t~1}^{~0}(x)~=~\omega_{t~0}^{~1}(x)&=& \frac{r_s}{2r^3}\sqrt{\frac{\Delta}{f}}\\\nonumber
\omega_{\phi~1}^{~0}(x)~=~\omega_{\phi~0}^{~1}(x)&=& -\frac{ar_s}{2r^3}\sqrt{\frac{\Delta}{f}}\\\nonumber
\omega_{r~3}^{~0}(x)~=~\omega_{r~0}^{~3}(x)&=& \frac{ar_s}{2r^2f\sqrt{\Delta}}\\\nonumber
\omega_{\theta~2}^{~1}(x)~=~-\omega_{\theta~1}^{~2}(x)&=& -\frac{\sqrt{\Delta}}{r}\\\nonumber
\omega_{t~3}^{~1}(x)~=~-\omega_{t~1}^{~3}(x)&=& -\frac{ar_s}{2r^3\sqrt{f}}\\\nonumber
\omega_{\phi~3}^{~1}(x)~=~-\omega_{\phi~1}^{~3}(x)&=& \frac{a^2r_s}{2r^3\sqrt{f}}-\sqrt{f}\\\nonumber
\end{eqnarray} 

\subsection{Geodesics}

The geodesics in a gravitational field can be obtained using the Lagrangian:
\begin{equation}
{\cal L}=\dot x^\mu \dot x^\nu g_{\mu\nu}(x)~
\end{equation}
and the associated Euler-Lagrange equations:
\begin{equation}
\frac{d}{d\tau}
\left( \frac{\partial {\cal L}}{\partial \dot x^\mu} \right)=0
\end{equation}
where:
\begin{equation}
\dot x^\mu=\frac{dx^\mu}{d\tau}=u^\mu(x)=\left(u^t(x),u^r(x),u^\theta(x),u^\phi(x) \right)
\end{equation}
normalized as:
\begin{equation}
u^\mu(x)u_\mu(x)=-1
\end{equation}
for massive test particles of unit mass ($m=1$).

In the equatorial plane ($\theta=\pi/2$), the geodesic equations for free falling massive particles are found to be:
\begin{eqnarray}
u^t(x)&=&\frac{ar_s}{r\Delta}\left( \frac{K}{\omega}-J \right)\\\nonumber
u^\phi(x)&=& \frac{ar_s}{r\Delta}\left( K +\frac{rfJ}{ar_s} \right)\\\nonumber
u^\theta(x)&=&0 \\\nonumber
\left(u^r(x) \right)^2&=&(K^2-1)+\frac{r_s}{r}+\frac{a^2(K^2-1)-J^2}{r^2}+\frac{r_s(J-aK)^2}{r^3}\nonumber
\end{eqnarray}
Because the Kerr metric tensor is independent of $t$ and $\phi$,  $K$ and $J$ are integration constants associated to the covariant components of the particle's 4-momentum that
are conserved along the geodesic:
\begin{eqnarray}
p_t&=&K\\\nonumber
p_\phi&=&J\nonumber
\end{eqnarray}
These two conserved quantities corespond to energy and angular momentum conservation, respectively \cite{carter}.

Because of the frame dragging due to the rotation
of the compact object, the geometry induces an 
angular velocity to a free falling particle, even if it has zero angular momentum ($J=0$):
\begin{equation}
\frac{d\phi}{dt}=\frac{u^\phi(x)}{u^t(x)}=\omega
\end{equation}
In the following section we will explore the effect of this frame dragging on qubit states.

Finally, it is important to recall that,
in contrast to orbits in Schwartzschild spacetime, non-equatorial
orbits in Kerr spacetime are not constrained to a plane \cite{hobson}. Then, in the sake of simplicity, we will restrict our discussion
to equatorial trajectories ($\theta=\pi/2$).

\section{Qubits in Kerr Spacetime}

We will limit our discussion of the effects of frame dragging in Kerr spacetime to equatorial radial falls and
equatorial circular orbits.

\subsection{Equatorial Radial Fall ($\theta=\pi/2$, $J=0$)}

A radial fall is a free falling particle with zero angular momentum ($J=0$)
and dropped from rest at $r\rightarrow\infty$ ($K=1$). 
The geodesics are given by:
\begin{eqnarray}
u^t(x)&=&\frac{ar_s}{\Delta r\omega}\\\nonumber
u^\phi(x)&=&\frac{ar_s}{\Delta r}\\\nonumber
u^\theta(x)&=&0\\\nonumber
(u^r(x))^2&=&\frac{r_s}{r}+\frac{a^2r_s}{r^3}\nonumber
\end{eqnarray}

We would like to use these geodesics to express the free falling motion of spin-$\frac{1}{2}$ particles. However, we 
have to recall that spin and curvature are coupled in a non-trivial manner \cite{wald}. 
As a consequence, the motion of spinning
particles, either classical or quantum, does not follow geodesics \cite{papa1, papa2, ubukhov, fenyk, fenyk2}. 
It is also known that the deviation from geodetic
motion is very small, except for the case of supermassive compact objects and/or ultra-relativistic 
test particles \cite{plyatsko1, silenko, plyatsko2, singh}.

However, in the two examples considered in this paper, 
the ${\cal O}(\hbar)$ non-geodesic motion induced by the coupling between the spin and the curvature can be safely
ignored.
Indeed, from the geodesic equations we can observe that, in the regime of interest, the geodetic motion is not ultra-relativistic: 
\begin{equation}
0\le\frac{r_s}{r}\le 1~~~~\Longrightarrow~~~~\left|\frac{dr}{dt}\right|\le0.4~~~~~~~~~~~~\left|r\frac{d\phi}{dt}\right|\le\frac{1}{3}
\end{equation}
Furthermore, let us consider the angular velocity $\omega_s$ produced by the non-geodetic motion that results from 
the spin-curvature coupling \cite{alsing2}.
If we require for this correction to be of about the same value as the maximum angular velocity due to frame dragging :
\begin{equation}
r\omega_{s}=\left|r\frac{v^\phi}{v^t}\right|\le\frac{1}{2r_sr}\left(1-\frac{r_s}{r}\right)\hbar;
~~~~~~~~r\omega_s\approx\frac{1}{3}   \Longrightarrow  r\approx 2r_s\approx 10^{-17} m
\end{equation}
which implies a circular orbit of femtometric radius around a supermassive black hole.

Using the geodesic equations we can calculate the 6 non-zero Lorentz transformations that describe the motion of the
particle in the local inertial frame:
\begin{eqnarray}
\lambda^0_{~1}(x)~=~\lambda^1_{~0}(x) &=&  \frac{ar_s^2}{2\Delta r^4}\left( a-\frac{1}{\omega} \right)\sqrt{\frac{\Delta}{f}}\\\nonumber
\lambda^0_{~3}(x)~=~\lambda^3_{~0}(x) &=& \frac{ar_s}{2r^2f\sqrt{\Delta}}\sqrt{\frac{r_s}{r}\left( 1+ \frac{a^2}{r^2} \right)}\\\nonumber
\lambda^1_{~3}(x)~=~-\lambda^3_{~1}(x) &=&  \frac{a^2r_s^2}{2\Delta r^4\sqrt{f}}\left( \frac{1}{\omega}-a\right) 
+\frac{ar_s\sqrt{f}}{\Delta r}
\nonumber
\end{eqnarray} 
These Lorentz transformations represent boosts on the 1 and 3 directions, and rotations over the 2-axis.

The associated infinitesimal Wigner rotation is found to be:
\begin{eqnarray}
\vartheta^1_{~3}(x) &=& \frac{a^2r_s^2}{2\Delta r^4\sqrt{f}}\left( \frac{1}{\omega}-a\right) +\frac{ar_s\sqrt{f}}{\Delta r} \\\nonumber
&& +\frac{ar_s^2}{2\Delta r^4(f+\sqrt{f})}\left( a^2f+r^2+\frac{ar_s}{\omega r}\right)
\nonumber
\end{eqnarray}
And the total angular rotation of the particle's spin, which is solely due to gravitational effects, is given by:
\begin{equation}
\Omega=\int_{\tau_i}^{\tau_f}\vartheta^1_{~3}(x) d\tau=\int_{r_i}^{r_f}\frac{\vartheta^1_{~3}(x)}{u^r(x)}dr
\end{equation}
where the integration takes place over the path of the free falling particle.

\begin{figure}[]
\label{fig:fig2}
\centering
\includegraphics[]{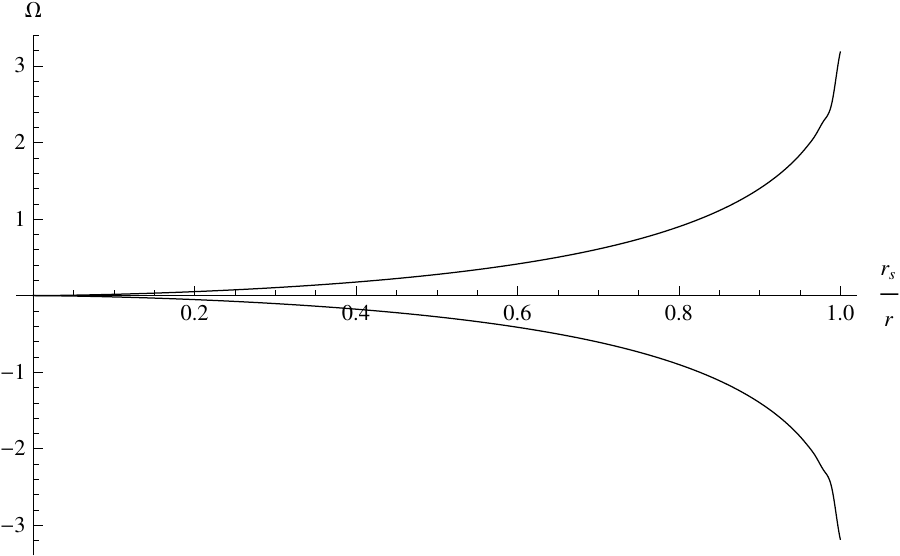}
\caption{Total spin rotation angle $\Omega$ induced by frame dragging on a free falling particle with zero 
angular momentum in the equatorial plane.
The two solid lines correspond to the upper and lower bounds of the angular momentum of the compact 
object ($a=0.5,-0.5$, respectively).}
\end{figure}

Figure 1 shows
the rotation angle $\Omega$ induced by frame dragging on a free falling particle with zero
angular momentum launched from rest at infinity. Thus, the integration over the radial coordinate
is taken from infinite to $r$. As previously discussed, the minimum value of $r$ is taken to be the
stationary limit surface  $S^+$, which takes the value of $r_s$ in the equatorial plane.

Furthermore, to guarantee the existence of event horizons and avoid naked singularities, the
ratio $a/r_s$ is only allowed to take values between $-0.5$ and $0.5$. 
When $a=0$ the spacetime geometry reduces to Schwarzschild and a radially falling particle
with zero angular momentum does not experience any rotation ($\Omega=0$). 
As the rotation of the compact body increases, so
does the Wigner rotation angle. That is, in this case the spin rotation angle $\Omega$ is completely due to frame dragging.
The two solid lines in Figure 2 correspond to the upper and lower bounds of the angular momentum of 
the compact object. 
The maximal rotation is reached at the stationary limit surface and has a value of 
$\Omega(S^+)\approx3.1828$.

\subsection{Equatorial Circular Orbits ($\theta=\pi/2$, $\dot x^r=0$)}

The geodesics for a circular orbit ($\dot x^r=0$) of constant radius $r$ in the equatorial plane ($\theta=\pi/2$) are given by:
\begin{eqnarray}
u^t(x)&=&\frac{ar_s}{\Delta r}\left(  \frac{K}{\omega}-J\right)\\\nonumber
u^\phi(x)&=&\frac{1}{\Delta}\left( \frac{aKr_s}{r} + fJ \right)\\\nonumber
u^\theta(x)&=&0 \\\nonumber
u^r(x)&=&0 \nonumber
\end{eqnarray}
and the constants $K$ and $J$ take the following values:
\begin{eqnarray}
K&=&\frac{1-\frac{r_s}{r}\mp a\sqrt{\frac{r_s}{2r^3}}}{\sqrt{1-\frac{3r_s}{2r}\mp 2a\sqrt{\frac{r_s}{2r^3}}}}  \\\nonumber
J&=&\mp\frac{1+\frac{a^2}{r^2}\pm 2a\sqrt{\frac{r_s}{2r^3}}}{\sqrt{1-\frac{3r_s}{2r}\mp 2a\sqrt{\frac{r_s}{2r^3}}}} ~\sqrt{\frac{rr_s}{2}} \nonumber
\end{eqnarray}
where the upper and lower signs correspond to counter-rotating and co-rotating circular orbits, respectively.
It can be observed that the values of $a$ an $r_s$ have a further restriction. Indeed,
from the equations for $K$ and $J$ we have the condition for circular orbits in Kerr metric:
\begin{equation}
1-\frac{3r_s}{2r}\mp 2a\sqrt{\frac{r_s}{2r^3}}>0~~\Longrightarrow~~ |a|<\left( 1-\frac{3r_s}{2r}\right)\sqrt{\frac{r^3}{2r_s}}
\end{equation}
which combined with the bound from Penrose's cosmic censorship hypothesis leads to the limiting values:
\begin{equation}
0\le\frac{r_s}{r}\le \frac{2}{3}
\end{equation}
As in the previous case, it can be observed that in the regime of interest, the motion is not ultra-relativistic. Indeed: 
\begin{equation}
\left|r\frac{d\phi}{dt}\right|\le 0.7
\end{equation}
And once more, the ${\cal O}(\hbar)$  non-geodesic motion induced by the coupling between the spin and the curvature can be safely
ignored.

The Lorentz transformations that describe the motion of the particle in the local inertial frame are:
\begin{eqnarray}
\lambda^0_{~1}(x)~=~\lambda^1_{~0}(x) &=&  \frac{ar_s^2}{2r^4\sqrt{\Delta f}}\left( 
J-\frac{K}{\omega}+aK+\frac{fJr}{r_s}
\right)\\\nonumber
\lambda^1_{~3}(x)~=~-\lambda^3_{~1}(x) &=&  \frac{a^2r_s^2}{2\Delta r^4\sqrt{f}}\left( \frac{K}{\omega}-J\right) \\\nonumber
&&~-\frac{ar_s}{\Delta r}\left( K+\frac{fJr}{ar_s} \right)\left( \frac{a^2r_s}{2r^3\sqrt{f}}-\sqrt{f}\right) 
\\\nonumber
\end{eqnarray}
These transformations correspond to a boost in the 1-direction and a rotation over the 2-direction.

The associated infinitesimal Wigner rotation is:
\begin{eqnarray} 
\vartheta^1_{~3}(x) &=&
\frac{ar_s^2\sqrt{f}}{2r^4\Delta (K+\sqrt{f})}\left( J-\frac{K}{\omega}+aK+\frac{fJr}{r_s}\right)\\\nonumber
&&~\times\left( -\frac{aK}{f}+aK+J \right) \\\nonumber
&&~+\frac{a^2r_s^2}{2\Delta r^4\sqrt{f}}\left( \frac{K}{\omega}-J\right) \\\nonumber
&&~-\frac{ar_s}{\Delta r}\left( K+\frac{fJr}{ar_s} \right)\left( \frac{a^2r_s}{2r^3\sqrt{f}}-\sqrt{f}\right) \nonumber
\end{eqnarray}
and therefore, the Wigner rotation per circular orbit is:
\begin{equation}
\Omega=\int_{\tau_1}^{\tau_2}\vartheta^1_{~3}(x) d\tau
=\int_0^{2\pi}\frac{\vartheta^1_{~3}(x)}{u^\phi(x)} d\phi
=2\pi \frac{\vartheta^1_{~3}}{u^\phi}
\end{equation}
because $\vartheta^1_{~3}$ and $u^\phi$ have fixed values, independent of the coordinates.

To analyze the gravitational effects on the spin of the orbiting particle, we need to substract the $2\pi$ due to the
trivial rotation of the particle around the compact object:
\begin{equation}
\Omega=2\pi-2\pi\delta\Omega~~~~\Longrightarrow~~~~\delta\Omega=1-\frac{\vartheta^1_{~3}}{u^\phi}
\end{equation}
where $\delta\Omega$ is the spin rotation angle (per orbit) entrely due to gravitational effects.

\begin{figure}[]
\label{fig:fig1}
\centering
\includegraphics[]{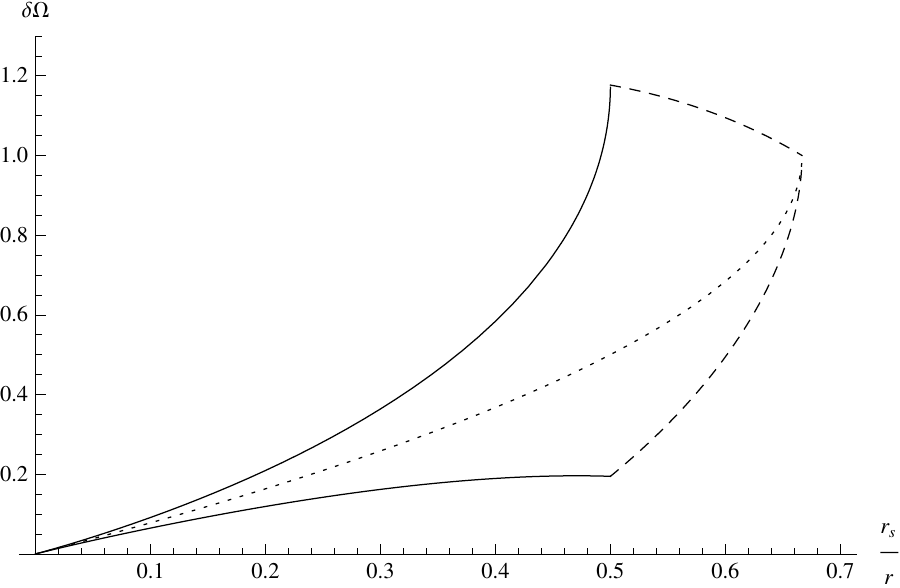}
\caption{Possible values for the gravitationally-induced spin rotation 
angle $\delta\Omega$ (per orbit) for a particle moving in an equatorial circular orbit in Kerr spacetime. The two solid lines correspond to
the upper and lower bounds imposed by the existence of event horizons. The two dashed lines correspond to the upper
and lower bounds imposed by the dynamics of circular orbits. The dotted line corresponds to circular orbits in Schwarzschild spacetime.}
\end{figure}

The possible values for the spin rotation angle (per orbit) due to gravity $\delta\Omega$ for a particle moving on an equatorial circular orbit in Kerr spacetime are shown in Figure 2. The two solid lines correspond to the upper and lower bounds imposed by the existence of event horizons $r_\pm$. That is, the upper and lower bounds corresponds to the maximal angular momentum
of the compact object:
\begin{equation}
a^{\pm}=\pm \frac{r_s}{2}
\end{equation}
Notice the dotted line in Figure 2, which represents the case where the Kerr metric reduces to the
Schwarzschild metric ($a=0$). 

The bounds on the angular momentum of the compact object imposed by the dynamics of the circular orbit further
restric the space of values for $\delta\Omega$. The two maximal values correspond to:
\begin{equation}
a^{\pm}_\circ=\pm \left( 1-\frac{3r_s}{2r}\right)\sqrt{\frac{r^3}{2r_s}}
\end{equation}
and are shown as dashed lines in Figure 2. These bounds generalize the well known gravitational effect that forbids 
circular orbits of arbitrarily small radius in Schwarzschild spacetimes.

The spin dynamic described by Figure 2 is expected. Even in the case of a non-rotating compact object ($a=0$), the
gravitational field affects the particle's spin by $\delta\Omega_0$. 
If the particle co-rotates with the compact object, $a>0$, then the spin rotation angle increases: 
$\delta\Omega_+>\delta\Omega_0$. If the particle counter-rotates with the compact object, $a<0$, and the
spin rotation angle decreases: $\delta\Omega_-<\delta\Omega_0$. 
However, notice that except for $r\rightarrow\infty$, there is no circular orbit which completely mitigates the
gravitational effects on the spin: $\delta\Omega>0$. Also, the spin rotation angle is bounded for all
circular orbits: $\delta\Omega<1.2$. The spin rotation is additive on completion of a single orbit. Thus, 
the spin rotation angle after $N$ 
orbits around the compact object is given by: $\delta\Omega_N=N\delta\Omega$.

\subsection{Qubit State Transformation}

Let us consider qubit states moving in Kerr spacetime in an equatorial radial fall with zero angular momentum or on an
equatorial circular orbit. The change to the $\sigma_z$ basis states is given by:
\begin{eqnarray}
D(W(x)) |0\rangle &=& e^{i\Omega\sigma_y/2}|0\rangle ~=~ \cos\frac{\Omega}{2}|0\rangle 
+ \sin\frac{\Omega}{2}|1\rangle \\\nonumber
D(W(x)) |1\rangle &=& e^{i\Omega\sigma_y/2}|1\rangle ~=~ \cos\frac{\Omega}{2}|1\rangle 
- \sin\frac{\Omega}{2}|0\rangle \nonumber
\end{eqnarray}
Then, if we consider a general qubit state expressed in the $\sigma_z$ basis:  
\begin{equation}
|\Psi\rangle = \alpha | 0 \rangle + \beta | 1 \rangle;~~~~~~~~~~|\alpha|^2+|\beta|^2=1
\end{equation}
it will transforms as:
\begin{equation}
D(W(x))|\Psi\rangle=\left( \alpha\cos\frac{\Omega}{2} - \beta\sin\frac{\Omega}{2} \right)|0\rangle
+  \left( \alpha\sin\frac{\Omega}{2} + \beta\cos\frac{\Omega}{2} \right)|1\rangle
\end{equation}
If the orbiting qubit is measured in the $\Psi$ basis, the probability of measuring the orthogonal state is:
\begin{equation}
\epsilon = 1-|\langle\Psi|D(W(x))|\Psi\rangle|^2=\sin^2\frac{\Omega}{2}
\end{equation}
Then, unless the unitary operation that inverts the gravitational Wigner rotation is applied to the orbiting state, $\epsilon\neq 0$ could lead to errors in 
some quantum communication protocols or in EPR-type experiments.  

For example, from Figure 2 we can observe that there are configurations of $(a,r_s,r)$ such that $\delta\Omega\approx1.0472$. In these cases 
the associated error is $\epsilon\approx0.5$. As a consequence, the channel capacity of a wide class of 
quantum channels will be zero.

Similar errors are incurred in the study of EPR experiments in Kerr spacetime. 
For instance, let us consider $({\cal T, S, Q, R})$ as a set of directions for the measurement of EPR states \cite{ueda}. 
Then, the
Bell's inequality will take the modified form:
\begin{equation}
\langle {\cal QS} \rangle + \langle {\cal RS} \rangle +  \langle {\cal RT} \rangle -  \langle {\cal QT} \rangle = 
2\sqrt{2} \cos^2\Omega
\end{equation} 
That is, the violation to Bell's inequality {\it appears} to be reduced by the presence of a Wigner rotation angle $\Omega\ne 0$.
However, it is important to note that this is not the case. Indeed, the Wigner rotation is a local unitary operation which
does not affect entanglement. The apparent decrease in the violation of Bell's inequality is due to the observer 
using an inadequate set of directions $({\cal T, S, Q, R})$. If the observer knows the exact orbital paths of the
particles, then he can derive a modified set of directions $({\cal T', S', Q', R'})$ that maximaly violate Bell's inequality.

\section{Conclusions}

In this paper we studied the effect of gravitational frame dragging on orbiting qubits. In particular we considered the
Kerr metric and the case of qubits in equatorial radial fall with zero angular momentum and in equatorial circular orbits.
We provided analytical equations and numerical simulations that describe the effect of gravity on a spin-$\frac{1}{2}$ particle.
Even though the spin rotations induced by gravity are small, they are comulative over the number of orbits. 
Also, as the resulting Wigner rotation is an unitary operation, the state of a particle can be restored by applying the inverse
operation. However, this is true only if the exact transformation is known. This may not be the case in some applications, for instance,
in satellite dynamics. As such, it is important to understand quantum information in the presence of a Kerr metric,
which
is the simplest exact solution to Einstein's field equations that describes the spacetime nearby to Earth.

\end{document}